A Data-Driven Machine Learning Approach for Electron-Molecule Ionization Cross Sections

A. L. Harris* and J. Nepomuceno
Physics Department, Illinois State University, Normal, IL, USA 61790

**Abstract**

Despite their importance in a wide variety of applications, the estimation of ionization cross sections for large molecules continues to present challenges for both experiment and theory. Machine learning algorithms have been shown to be an effective mechanism for estimating cross section data for atomic targets and a select number of molecular targets. We present an efficient machine learning model for predicting ionization cross sections for a broad array of molecular targets. Our model is a 3-layer neural network that is trained using published experimental datasets. There is minimal input to the network, making it widely applicable. We show that with training on as few as 10 molecular datasets, the network is able to predict the experimental cross sections of additional molecules with an accuracy similar to experimental uncertainties in existing data. As the number of training molecular datasets increased, the network's predictions became more accurate and, in the worst case, were within 30% of accepted experimental values. In many cases, predictions were within 10% of accepted values. Using a network trained on datasets for 25 different molecules, we present predictions for an additional 27 molecules, including alkanes, alkenes, molecules with ring structures, and DNA nucleotide bases.

## 1. Introduction

Atomic and molecular cross sections play a pivotal role in many areas of applied physics, including plasma physics, biophysics, and astrophysics. In these fields, cross sections are essential fundamental inputs for modeling complex problems. The success of models within these fields to understand fundamental physical processes relies, at least in part, on the accuracy and availability of the scattering cross sections. Often, cross sections are required over a wide range of energies, target species, and collision processes. Databases such as the NIST Electron Elastic-Scattering Cross-Section Database [1], LxCat [2], BEAMDB [3], and others [4] have begun to address the need for large amounts of cross section data by compiling available experimental and theoretical datasets into openly accessible repositories. There are also computational methods that have proven to be reliable in predicting cross sections in various energy regimes, and many of these are being made publicly available through resources such as the Atomic, Molecular, and Optical Science Gateway [5]. Despite the increasing availability of cross section databases, in many instances, the necessary data remains inaccessible experimentally or too computationally demanding for ab initio theory. It is therefore impractical to rely exclusively on experiment or computation to obtain all of the needed data.

Machine learning (ML) algorithms have proven to be effective tools in many areas of physics. These algorithms utilize data from existing experiment and/or simulation to train a model that is able to accurately predict the data for unknown systems. The use of ML algorithms in the physical sciences has exploded in recent years and is becoming commonplace in many areas of physics, such as high energy physics [6,7], quantum many body problems [8], quantum computing [9], molecular chemistry and material science [10], and countless others. However, these techniques have seen only limited use in atomic and molecular collision physics. Existing applications in collision physics can be sorted into two broad categories that predict atomic and molecular cross sections: (1) training a ML model using measured or calculated cross sections or (2) using a ML model to solve the inverse swarm problem and predict the cross sections.

*alharri@ilstu.edu

In the first category, one of the earliest applications of ML techniques for atomic and molecular cross sections was implemented by El-Bakry and El-Bakry in which a neural network was trained on experimental data for the total cross sections for electron and positron impact collisions with sodium and potassium [11], and their predictions showed good agreement with measured data. Since then, several groups have developed various implementations of ML techniques. Harris and Darsey trained a neural network to predict proton-impact ionization double differential cross sections for atoms and molecules [12] and demonstrated the method's success with a limited number of inputs and training sets. El-Bakry et al. developed a model that used the gradient tree boosting method to predict total electron and positron scattering cross sections for alkali atoms [13], demonstrating that ML techniques other than neural networks are successful at predicting collision cross sections. Zhong trained a support vector machine algorithm that was used to predict larger molecule cross sections from the theoretical Binary Encounter Bethe (BEB) cross sections of small molecules [14]. While these results showed good agreement between the ML model and the BEB model, this approach was limited because it relied on the approximate calculation of smaller molecule cross sections as inputs. Effectively, the technique trained a ML algorithm to replace the BEB model for large molecules, which was limited to the accuracy of the theoretical model used to create the input training data. Amaral and Mohallem applied the sure independence screening and sparsifying operator method to classify bound and unbound systems in low energy positron-molecule scattering and made predictions of possible new bound systems [15]. This is one of the few ML applications in atomic and molecule collisions that uses the algorithm for classification of a physical process involved in scattering, rather than the prediction of cross sections. Jasinksi et al. used Bayesian ML to predict cross sections for inelastic scattering of two diatomic molecules [16]. Their results demonstrated that these types of algorithms can be used to improve approximate calculations using only a handful of rigorous calculations.

In the second category, known cross section data from available databases and the literature are combined with numerical solutions of the Boltzmann equation to solve the inverse swarm problem using ML techniques. The cross section data is then refined to provide complete and consistent datasets [17–19]. Stokes et al. created a complete and self-consistent set of cross section data for quasielastic momentum transfer cross sections, dissociative electron attachment cross sections, and neutral dissociation cross sections for nitric oxide [18]. They showed that the predicted cross section dataset has improved agreement with experimental data. Jetly et al. compared different machine learning algorithms, such as artificial neural networks, convolutional neural networks, and a densely connected convolutional network to predict cross sections from swarm data [17]. They demonstrated that the dense network approach gave more accurate predictions than the neural network techniques. Stokes et al. determined the elastic momentum transfer and ionization cross sections for helium and argon [19], and showed agreement with experiment within 4%.

In all of the above examples, the ML algorithms were able to successfully predict cross section data. However, one of the challenges with ML approaches is that they often require a large amount of data for training. In general, the more training data that is available to train the model, the better the network will predict the desired unknown data. However, it is exactly the lack of available data that motivates the use of ML algorithms for collision physics in the first place. Thus, the development of a ML model that can make accurate predictions with relatively limited amounts of data is therefore desirable and one of the goals of the work presented here. In this work, we focus on the prediction of total ionization cross sections for molecules of any size and a wide array of shapes and chemical compositions. We test the effectiveness of the algorithm with respect to the number of training datasets and input parameters.

Molecular cross sections were chosen because they present particular challenges to theory and experiment. Compared to atomic targets, molecular targets have a much more complex electronic and nuclear structure. From the theoretical perspective, multi-electron effects and orientation effects both influence the ionization cross sections, but remain difficult to model. Experimentally, molecular targets may be difficult to produce, have high reactivity, and can be expensive. All of these factors combined lead to the limited availability of molecular ionization cross sections.

To help fill these gaps, we have developed a ML-based model that uses a relatively small number of available experimental cross sections for molecular targets to train a neural network. Because experimental data is used for training, our algorithms are independent of any theoretical limitations inherent in theoretical models. Once trained, the network can make predictions for additional molecules. We show that even with limited training input, the network is able to successfully predict molecular ionization cross sections to within 30% in the worst case, and often to within less than 10%. We also provide a database of predicted molecular ionization cross sections for 27 additional target molecules. With the goal of making cross section predictions easily available to the community, we have made our training algorithm and the fully trained network (ready to make predictions) available on Figshare [20].

## 2. Methods

The ML algorithm that we use for predicting molecular ionization cross sections involves several steps. The first step is the identification and processing of training data. Step 2 is the creation of the network for training. Step 3 is the training of the network, and step 4 is the application of the trained network to predict new data. Each of these steps is described in detail in the sections below and all data sets and codes are available on Figshare [20].

### 2.1 Training Data

All training datasets used here were taken from published literature (see Table 1). We identified experimental data sets for 25 different molecules in the form of electron-impact total ionization cross sections as a function of projectile energy. In some cases, multiple data sets were available for a given molecular target, in which case, we selected the datasets that limited the number of independent sources from which the data was taken, while simultaneously giving preference to newer datasets. Each dataset was converted to units of $a_0^2$ for the cross section values and electron volts for the electron projectile energies. Because the energy range and energy grid of the measurements were not consistent between published data sets, we selected only data within an energy range common to all data sets (i.e. 25 to 100 eV). Each dataset was then linearly interpolated to a fixed grid of 101 energies.

### 2.2 Creating the Network

To help with terminology, we introduce some definitions. We define a *network* as the configuration of the number of nodes, layers, and their connectivity. All of our networks were 3-layer feed-forward neural networks, consisting of an input layer, one hidden layer, and an output layer (see Fig. 1). The networks were fully connected, such that every node in one layer is connected to every node in the adjacent layers. The input to a given node is a weighted sum of the values passed from each of the upstream nodes plus a bias. From this input, the output of the node is calculated using a logistic sigmoid activation function. This output is then passed downstream to the nodes in the next layer, where it is processed as the input. We define a network *trial* as the optimized weights and biases for a network that was trained with a specific set of training data.

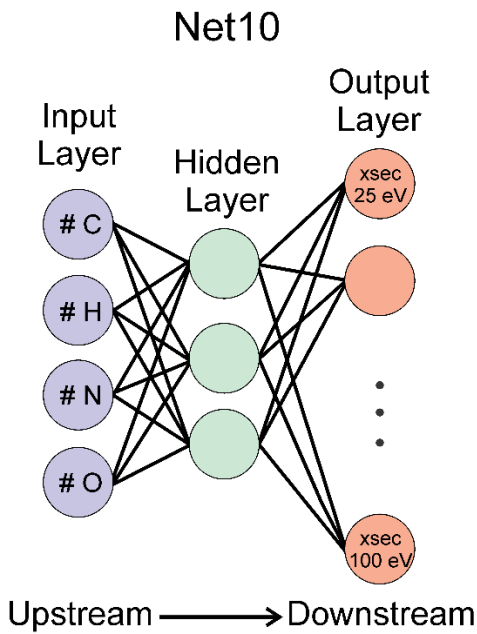

Figure 1 Diagram of network configuration for Net10. The network has 4 input nodes, corresponding to the number of carbon, hydrogen, nitrogen, and oxygen atoms in a molecule; 3 hidden nodes, which is equal to 1/3 of the number of training sets; and 101 output nodes, corresponding to the ionization cross section (xsec) at each energy ranging from 25 eV to 100 eV. The network is fully connected with every node in a layer connected to every node in the adjacent layers.

The following are examples of different networks and network trials. Net10 has 4 input nodes, 3 hidden nodes, and 101 output nodes. In contrast, Net15 has 4 input nodes, 5 hidden nodes, and 101 output nodes. We define Net10 and Net15 as different networks. Net15 Trial 1 are the optimized weights and biases for Net15 that was trained on a given set of 15 input training molecule datasets. Net15 Trial 2 are the optimized weights and biases for Net15 trained on a different set of 15 input training molecule datasets. The 15 training sets used in Net15 Trial 1 are different than the 15 training sets used in Net15 Trial 2.

Many different networks and trials were created for testing the effectiveness of our machine learning algorithm. In all networks, the number of nodes in the hidden layer was equal to one third of the number of training sets [21]. Because the number of training sets could vary, this resulted in different networks depending on the number of training sets. All networks had 101 nodes in the output layer, corresponding to the value of the cross section at each of the 101 projectile energies.

In some networks, the input layer consisted of 4 nodes, each corresponding to the number of carbon, hydrogen, nitrogen, and oxygen atoms in the molecule (i.e. the molecule's chemical formula). In other networks, the input layer consisted of 5 nodes, with 4 of the nodes corresponding to the number of carbon, hydrogen, nitrogen, and oxygen atoms in the molecule and a fifth node corresponding to the ionization potential of the molecule in eV. The input data were collectively normalized to a range of values between 0.05 and 0.95 to approximately match the range of the logistic sigmoid function. The biases and weights were initialized with pseudorandom values, which were optimized through backpropagation during the training phase.

### 2.3 Training the ML Model

For evaluation and testing of the ML models, the 25 cross section datasets were randomly sorted into two groups – the training group and the testing group. The training group contained 20 molecules (80%) and the testing group contained the remaining 5 molecules (20%). A list of these molecules, their chemical formula, their ionization potential, and a reference to the original data is provided in Table 1. A plot of the cross sections is shown in Fig. 2. Any training of the network used only the molecules in the training group. The testing group was set aside to provide a comparison of the network's prediction with data that the network had never seen. Two different partitions of the 25 datasets were made, partition 1 and partition 2, so that results could be evaluated for different testing datasets. In Fig. 2, the cross sections for the partition 1 testing dataset are shown as solid red data points and the cross sections for the partition 2 testing dataset are shown as solid blue data points. For partition 1, all datasets except those shown as solid red data points were included in the testing set, while for partition 2 all datasets except those shown as solid blue data points were included in the testing dataset.

During training, the input parameters were fed into the network for a given molecule. These values were propagated through the network and then a prediction was made for the cross sections (i.e. the output). This prediction was compared to the known cross section values, and the weights and biases of the connections between nodes were adjusted to optimize the agreement between the network's prediction and the known cross section values. This process was repeated with each of the molecules from the training set. After cycling through all molecules in the training set, the process repeated, starting again with the first molecule. Each cycle through all of the training datasets is called an epoch, and we trained each network for 400,000 epochs. Convergence testing was done with a few trials training for 200,000 epochs and 1,000,000 epochs. No difference was observed between predictions made with 400,000 and 1,000,000 epochs, but differences were found when only 200,000 epochs were used. Thus, we used 400,000 epochs for all network training.

All codes were written and executed in Mathematica version 12.1 [22], and the network was created with the NetChain command. Training of the network was conducted using the function NetTrain with the default settings. Details are available in the open source code at Figshare [20].

### 2.4 Predictions

Once the network was trained, it was given new inputs that were propagated through the network to make a prediction. For evaluation purposes, the network was given inputs corresponding to the 5 testing datasets that were excluded from training. Then, the network's prediction was compared with the known, published experimental data. Six networks were used to make predictions, three that did not include the ionization potential as an input parameter and three that did include the ionization potential as an input parameter. Thus, the effect of the ionization potential as an additional input on the accuracy of the networks' predictions could be tested. In each case, the three networks corresponded to different numbers of training datasets so that the effect of the number of training molecules on the networks' predictions could be examined.

Lastly, 10 trials of each network were performed to evaluate how a network's predictions depended on the identity of the training molecules.

| IUPAC Molecule Name | Chemical Formula | Ionization Potential (eV) | Experimental Data Reference |
|---|---|---|---|
| Ethanal** | $C_2H_4O$ | 10.23 | [23] |
| Propanal** | $C_3H_6O$ | 9.96 | [23] |
| Butanal | $C_4H_8O$ | 9.82 | [23] |
| 2-Methylpropanal* | $C_4H_8O$ | 9.71 | [23] |
| Ethoxyethane | $C_4H_{10}O$ | 9.51 | [23] |
| Propoxypropane | $C_6H_{14}O$ | 9.3 | [23] |
| 2-Isopropxypropane | $C_6H_{14}O$ | 9.2 | [23] |
| Propanone* | $C_3H_6O$ | 9.7 | [23] |
| Butanone | $C_4H_8O$ | 9.52 | [23] |
| Pentan-2-one | $C_5H_{10}O$ | 9.38 | [23] |
| Pentan-3-one | $C_5H_{10}O$ | 9.31 | [23] |
| 3-Methylbutan-2-one** | $C_5H_{10}O$ | 9.31 | [23] |
| Hexan-3-one* | $C_6H_{12}O$ | 9.3 | [23] |
| Hexan-2-one | $C_6H_{12}O$ | 9.35 | [23] |
| 3,3-Dimethylbutan-2-one* | $C_6H_{12}O$ | 9.14 | [23] |
| 3-Methylpentan-2-one | $C_6H_{12}O$ | 9.2 | [23] |
| 4-Methylpentan-2-one | $C_6H_{12}O$ | 9.3 | [23] |
| Molecular Hydrogen | $H_2$ | 15.43 | [24] |
| Molecular Nitrogen** | $N_2$ | 15.58 | [24] |
| Carbon Monoxide | $CO$ | 14.01 | [24] |
| Nitric Oxide | $NO$ | 9.26 | [24] |
| Molecular Oxygen | $O_2$ | 12.07 | [24] |
| Methanol* | $CH_4O$ | 10.84 | [25] |
| Ethanol** | $C_2H_6O$ | 10.48 | [25] |
| Water | $H_2O$ | 12.62 | [26] |

Table 1 List of molecules whose ionization cross sections used as training datasets. Molecules with a single asterisk (*) were excluded from the training dataset in partition 1 and made up the testing dataset for partition 1. Molecules with two asterisks (**) were excluded from the training dataset in partition 2 and made up the testing dataset for partition 2. The ionization potential values came from *[27]*.

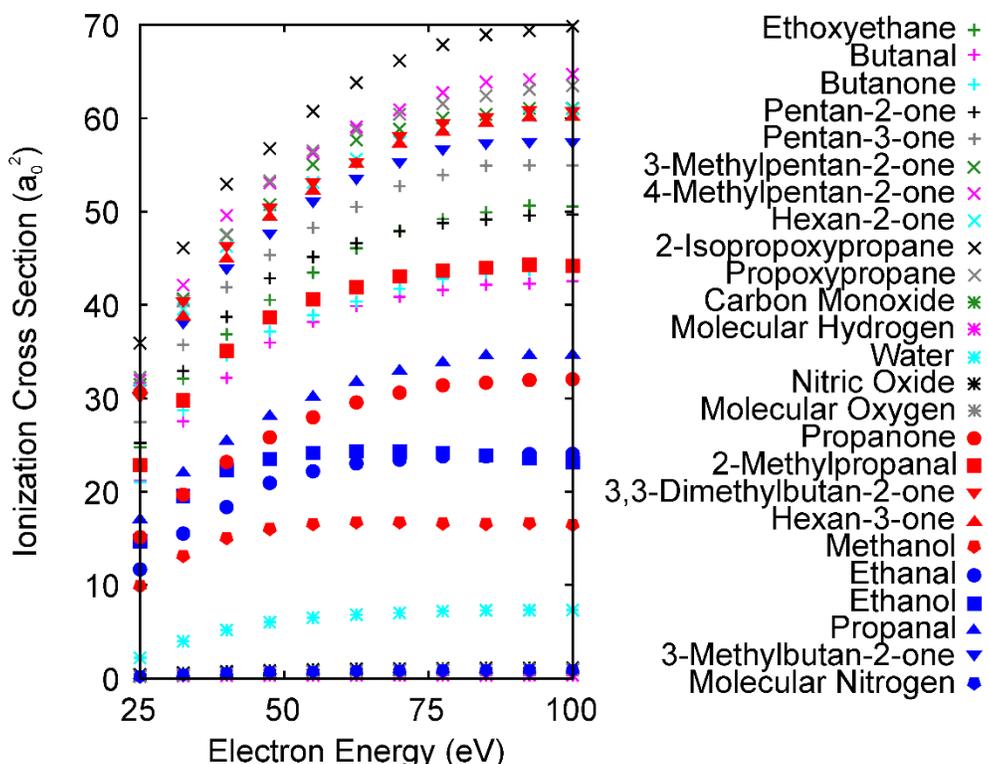

Figure 2 Interpolated experimental ionization cross sections used for training in partitions 1 and 2 (only 11 of the 101 data points are shown for clarity). See text for a description of how published datasets were processed for use in the algorithm. Solid red (blue) data points denote datasets excluded from training (and included in the testing datasets) in partition 1 (partition 2). References for the experimental datasets can be found in Table 1.

## 3. Results

To determine the effect of the number of training sets and the identity of the training molecules on the algorithm's ability to make accurate predictions, we compared predicted cross sections for three networks. Net10 had 4 input nodes, 3 hidden nodes, and 101 output nodes, and was trained on cross sections for 10 different molecules that were randomly selected from the 20 training datasets. Net15 had 4 input nodes, 5 hidden nodes, and 101 output nodes, and was trained on cross sections for 15 different molecules that were randomly selected from the 20 training datasets. Net20 had 4 input nodes, 6 hidden nodes, and 101 output nodes and was trained on all 20 training datasets. For Net10 and Net15, 10 trials were conducted, each with a different random selection of molecules from the training datasets. From these 10 trials, an average cross section and standard deviation were calculated.

Figure 2 shows the predicted cross sections from Net10, Net15, and Net20. The 5 testing molecules were 2-Methylpropanal, Propanone, Hexan-3-one, 3,3-Dimethylbutan-2-one, and Methanol. Each trial's predicted cross section is shown as a dashed-dotted line for Net10 and Net15, and the average of the 10 trials is shown as a solid blue line. For Net20, only 1 trial was calculated because all 20 training molecules were included in the training set, and this prediction is shown as a solid blue line. The light blue shaded region represents the standard deviation of the

10 trials from the average for Net10 and Net15. The experimental data for the 5 testing molecules is shown in Fig. 2 as open circles.

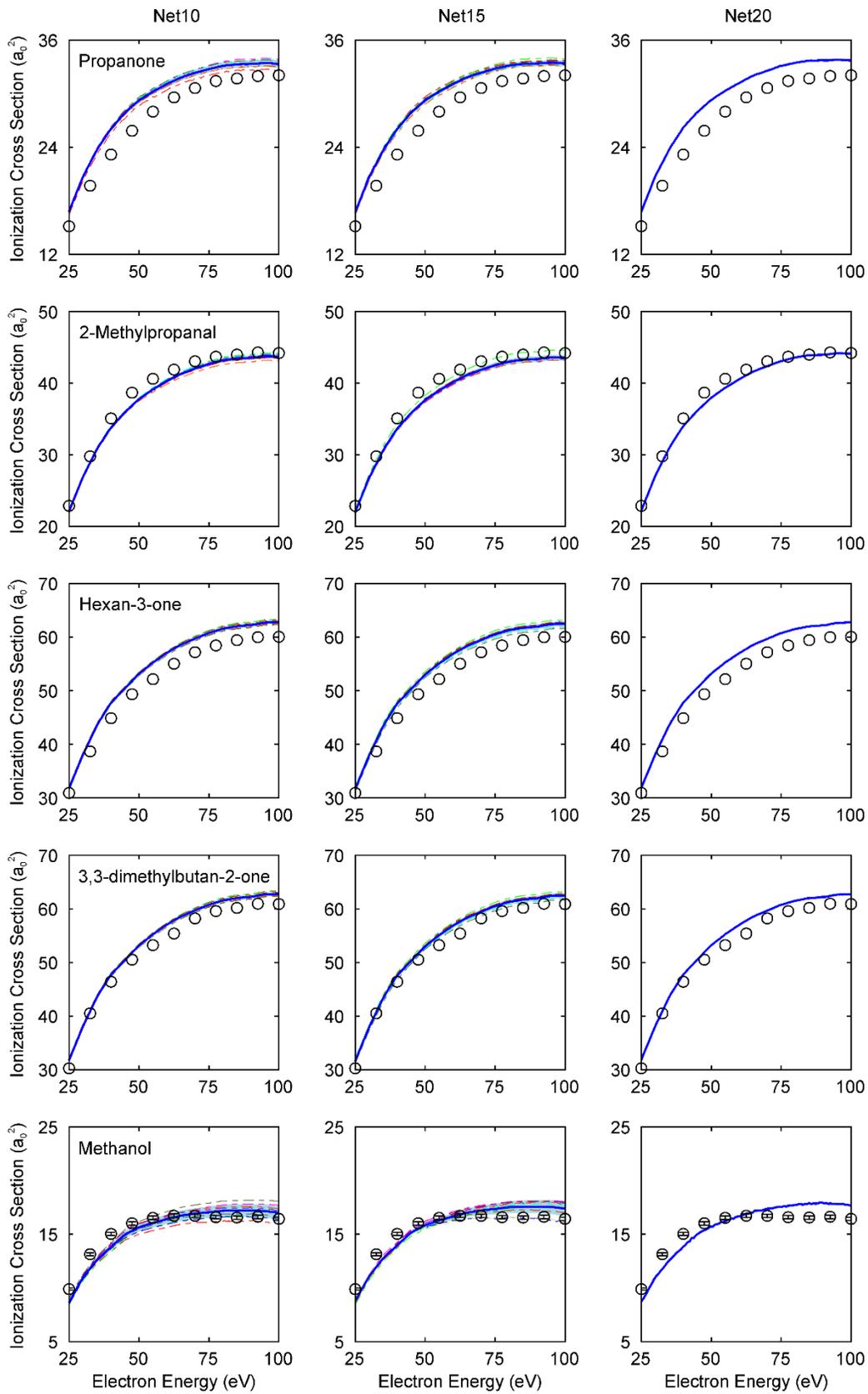

Figure 3 Predicted ionization cross sections as a function of projectile energy using three different networks (Net10 – column 1, Net15 – column 2, Net20 – column 3) with training data taken from partition 1.  The ionization potential was not included as an input parameter.  The testing molecules are labeled in the figure for each row.  Dash-dotted lines (columns 1 and 2) are predictions from the 10 different trials of the networks.  The solid blue line is the average of the predictions of the 10 trials, and the shaded blue area is the standard deviation of the predictions of the 10 trials.  Interpolated experimental data are open circles (see Table 1 and Fig. 2 for references and details).

The algorithm's predictions show the most variation among the trials when only 10 datasets were used for training (i.e. Net10).  The deviation of the prediction varied with projectile energy, and in the worst case (methanol at 100 eV), the average of the trial predictions is 17.0 a.u. with a standard deviation of 0.63 a.u.  When 15 training datasets were used (i.e. Net15), less variation among the trials was observed, indicating that with more training molecules, the identity of the training molecules had a decreased importance.

For all 5 testing molecules, the use of 20 training datasets yielded excellent predictions of the cross sections.  For all of the molecules, the maximum percent difference between the Net20 predictions and the known experimental values was less than 14% (Propanone at 35 eV), and most percent differences were in the single digits.  The maximum percent differences between the Net20 prediction and the experimental data is given in Table 2 for each of the molecules.  It is interesting to note that the prediction of Net20 was nearly identical to the averages of the predictions from the trials using Net10 and Net15.  This indicates that the variation in prediction caused by different molecules in the training dataset was averaged out when multiple trials were conducted.  For practical applications of our algorithm, one would ideally include all available datasets for training, thus eliminating the possibility of averaging over trials.  However, based on the results shown in Fig. 2, accurate predictions can be achieved with as few as 10 training sets.

For Net10, Net15, and Net20, 4 inputs were used, corresponding to the number of carbon, hydrogen, nitrogen, and oxygen atoms in the molecule.  The good agreement between the network predictions and experimental data for the 5 testing molecules in Fig. 2 demonstrated that the molecular ionization cross sections were strongly correlated with the chemical formula of the molecule.  However, some discrepancies remain, particularly for the smaller molecules, such as methanol and propanone.  Also, the agreement between prediction and experiment becomes slightly worse as the projectile energy increases.  To test whether these discrepancies could be resolved with an additional input parameter, specifically the molecule's ionization potential, we created 3 new networks that included the ionization potential as an input.  We refer to these networks as Net10Ip, Net15Ip, and Net20Ip.  Their network configuration is identical to Net10, Net15, and Net20, except that they have an additional input node corresponding to the ionization potential. We repeated the process described above for these 3 new networks.  For Net10Ip and Net15Ip, 10 trials were performed with different randomly selected training datasets and the average and standard deviation of the predictions was calculated.

| Partition 1 | | |
| --- | --- | --- |
| Molecule | Net20 Maximum % Difference | Net20Ip Maximum % Difference |
| Propanone | 14 | 16 |
| 2-Methylpropanal | 4 | 5 |
| Hexan-3-one | 7 | 7 |
| 3,3-Dimethylbutan-2-one | 5 | 6 |
| Methanol | 12 | 9 |

| Partition 2 | | |
|---|---|---|
| Molecule | Net20 Maximum % Difference | Net20Ip Maximum % Difference |
| Ethanal | 13 | 14 |
| Ethanol | 26 | 23 |
| Propanal | 6 | 8 |
| 3-Methylbutan-2-one | 13 | 12 |
| Molecular Nitrogen | 30 | 1940 |

Table 2 Maximum percent difference between the Net20 and Net20Ip cross section predictions and the experimental data for the datasets in the testing groups.

Figure 3 shows the averages and standard deviations for Net10, Net15, and Net 20 in blue and for Net10Ip, Net15Ip, and Net20Ip in red. For clarity, no individual trial predictions are shown in Fig. 3. Surprisingly, the inclusion of the ionization potential as an additional input parameter for the networks yielded greater variation among the trials and generally worse agreement with experiment. The agreement between predictions and experiment only improved for Methanol with Net20Ip and Hexan-3-one with all networks when the ionization potential was included. This indicated that there was not a strong or useful correlation between ionization potential and cross section. In addition, because the standard deviation was larger for the predictions of Net10Ip and Net15Ip than for Net10 and Net15, more training data sets were required to achieve accurate predictions when the ionization potential was used as an input parameter.

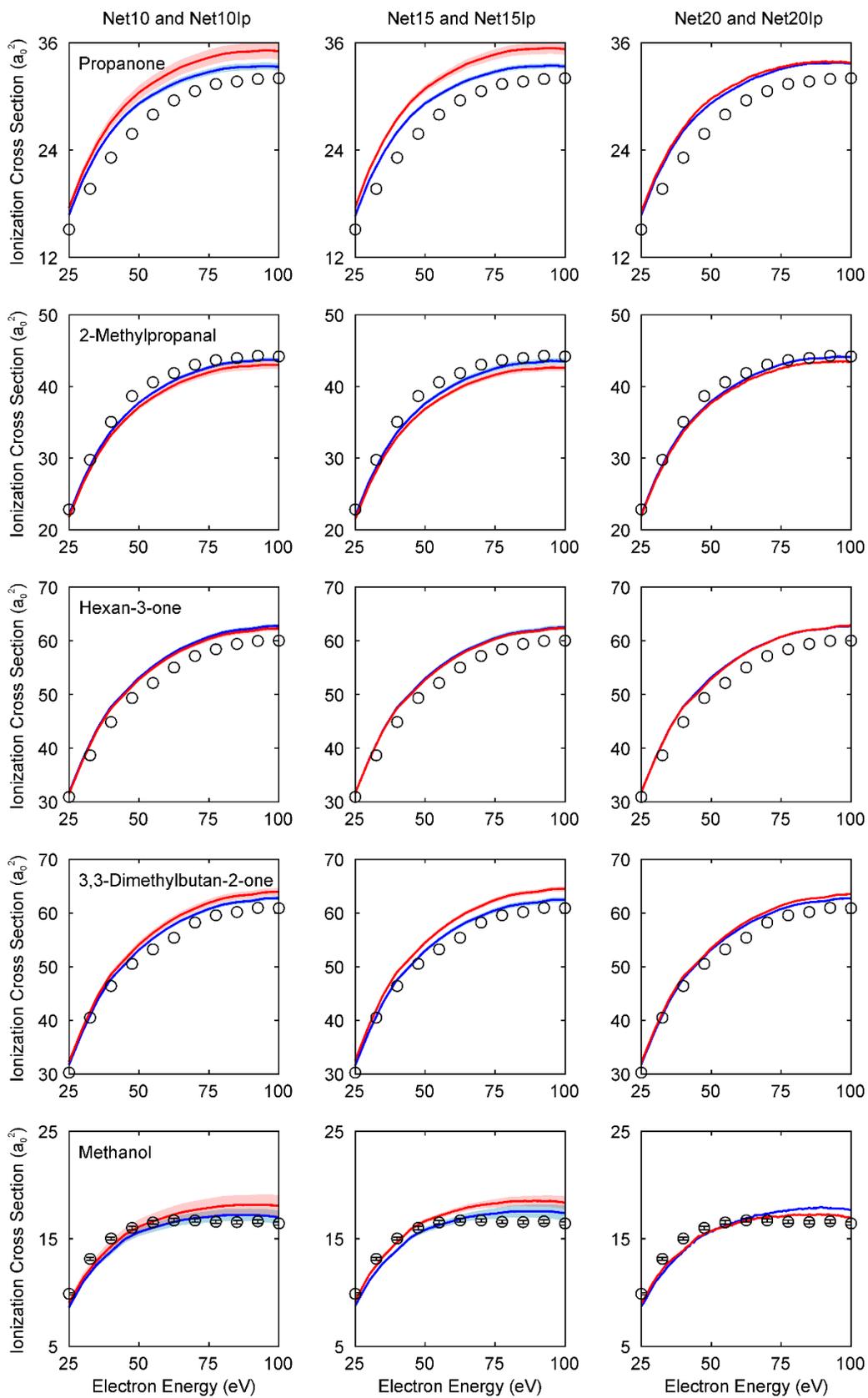

Figure 4 Same as Fig. 2, but includes ionization cross section predictions from Net10Ip, Net15Ip, and Net20Ip in which the ionization potential of each molecule was included as an additional input parameter to the network. The red solid line is the average of the predictions for 10 trials for networks that include the ionization potential as an input parameter and the red shaded area is the standard deviation for these predictions.

To further test the role of the training molecule identities on the algorithm's ability to accurately predict cross sections, we repeated the above training and predictions using a different partition of the 25 datasets into 20 training datasets and 5 testing datasets. In partition 2, the 5 testing datasets were for Ethanal, Ethanol, Propanal, 3-Methylbutan-2-one, Molecular Nitrogen. Figure 5 shows the predicted cross sections from Net10, Net15, Net20, Net10Ip, Net15Ip, and Net20Ip for the 5 testing datasets in partition 2. As before, 10 trials were performed for Net10, Net15, Net10Ip, and Net15Ip and an average and standard deviation of the cross sections was calculated.

Some similar trends were observed in partition 2 compared to partition 1. The variation among trials was larger for the networks that included the ionization potential as an input, and this variation decreased when more training datasets were used. Agreement between the algorithm's predictions was worse for some molecules when the ionization potential was included (Ethanal, Propanal, Molecular Nitrogen) and better for others (Ethanol, 3-Methylbutan-2-one), with the best agreement at low energy. Notably, for Molecular Nitrogen, only the networks that did not include the ionization potential were able to reasonably predict the experimental data. This is likely due to a combination of factors. First, only 1 training dataset for a molecule with a nitrogen atom was included in partition 2 (Nitric Oxide). Second, the cross sections for diatomic molecules are at least an order of magnitude smaller than those for the larger hydrocarbons. Third, the ionization potential of Molecular Nitrogen is larger than most of the other molecules included in the training dataset (see Table 1). All of these factors likely combined to make it more difficult for the algorithm to predict the cross sections for diatomic nitrogen. We note that despite the much smaller magnitude of the diatomic nitrogen cross sections, the networks without the ionization potential do a remarkable job of accurately predicting these cross sections.

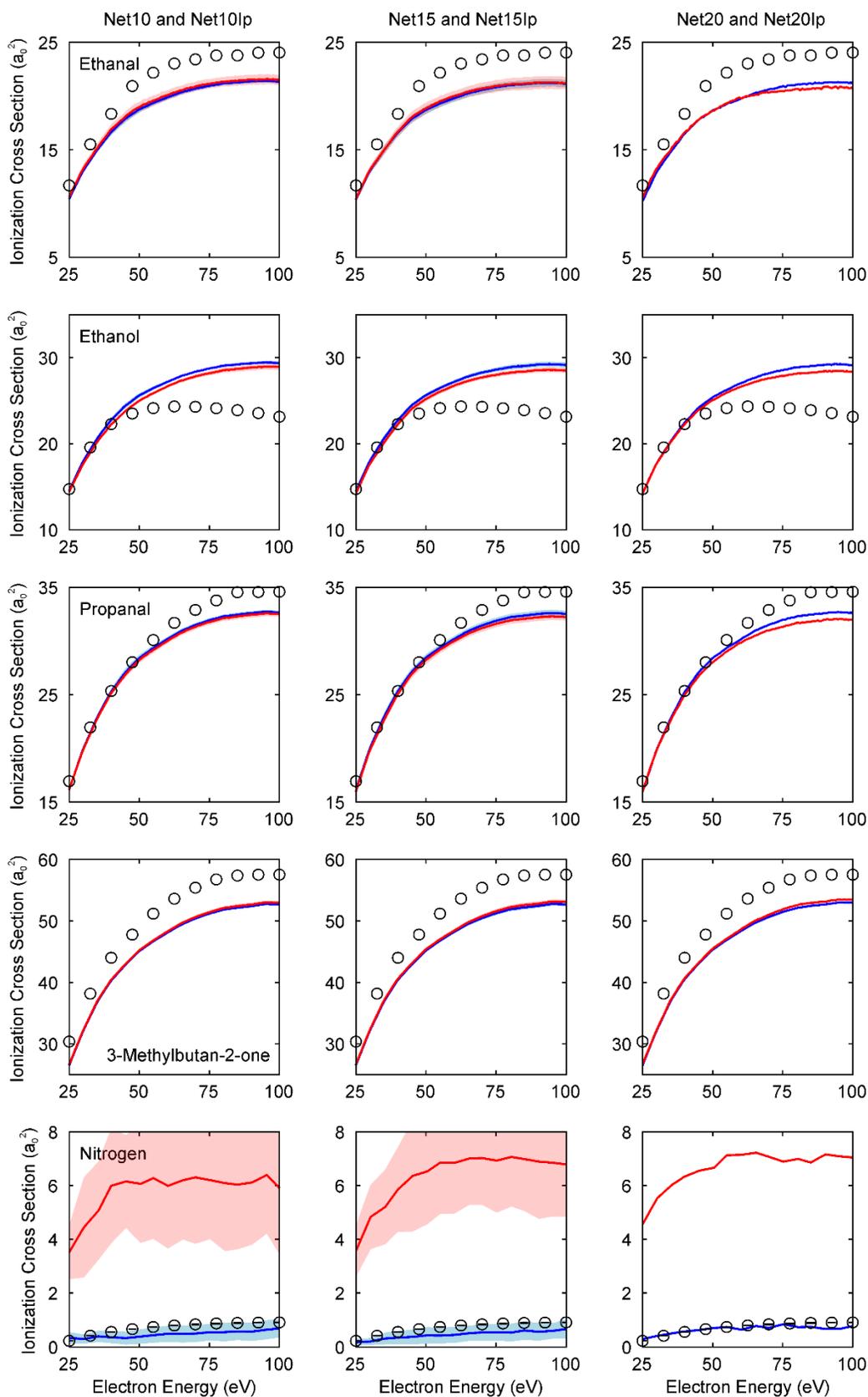

Figure 5 Same as Fig. 4, but using the training and testing datasets from partition 2.

Overall, for the molecules in partition 2, the agreement between the algorithm's predictions and the experimental data is generally worse than for partition 1. In particular, the predictions for the testing molecules in partition 2 underestimate the high energy cross sections for Ethanal, Propanal, and 3-Methylbutan-2-one, while the prediction for Ethanol overestimates the high energy cross section. The overestimation for ethanol is likely due to the different shape of this experimental cross section relative to the molecules included in the training set. Figure 1 shows that only the Ethanol experimental cross section has a maximum (around 60 eV) and then a decreasing cross section for higher energy (greater than 60 eV). Because the networks in partition 2 had no training data with a local maximum, it was not possible for the prediction to exhibit a local maximum. One possible explanation for the underestimation of the 3-Methylbutan-2-one and Propanal cross sections is that the training datasets included cross sections for isoforms of these molecules, which have smaller cross sections, thus leading to an underestimation in the prediction. In the case of Ethanal, the most similar training molecule was Methanol, which has a smaller cross section and likely led to the Ethanal prediction underestimating the experimental data.

The predictions from partitions 1 and 2 imply that while the identity of training molecules within a given partition does not significantly alter the network's prediction, the partition of the training datasets can influence the network's prediction accuracy. We note, however, that the partitioning of datasets was done purely for testing purposes to evaluate the algorithm's accuracy in predicting cross sections. Any applications of the algorithm would use all available datasets. Based on the worst case of the Net20 predictions (partition 2, Molecular Nitrogen), the algorithm very accurately predicted the cross section for energies less than 40 eV, and at 100 eV differed from the experimental data by 30%. Because the algorithm's predictions improve with more training datasets, we expect that predictions from a Net25 network including all 25 available datasets would have an error of less than 30%. This is acceptable and similar to experimental datasets, which can have an error of around 20% [25,28,29].

Given the success of our algorithm for predicting molecular ionization cross sections, we present in Fig. 6 predictions for additional molecules of interest. To produce these predictions, the network was trained on all 25 experimental datasets listed in Table 1 for 400,000 epochs, and the ionization potential was not included as an input parameter. The trained network was then given input values corresponding to each new molecule's chemical formula. The predictions for several classes of molecules are shown in Fig. 6, including alkanes, alkenes, molecules with a ring structure, and DNA nucleotide bases. The chemical formulae of the molecules are listed in Table 3.

| IUPAC Molecule Name | Chemical Formula |
|---|---|
| Methane | $CH_4$ |
| Ethane | $C_2H_6$ |
| Propane | $C_3H_8$ |
| Butane | $C_4H_{10}$ |
| Pentane | $C_5H_{12}$ |
| Hexane | $C_6H_{14}$ |
| Heptane | $C_7H_{16}$ |
| Octane | $C_8H_{18}$ |
| Ethene | $C_2H_4$ |
| Prop-1-ene | $C_3H_6$ |
| But-1-ene | $C_4H_8$ |
| Pent-1-ene | $C_5H_{12}$ |
| Hex-1-ene | $C_6H_{14}$ |
| Hept-1-ene | $C_7H_{16}$ |
| Oct-1-ene | $C_8H_{18}$ |
| Benzene | $C_6H_6$ |
| Cyclopropane | $C_3H_6$ |
| Cyclobutane | $C_4H_8$ |
| Cyclopentane | $C_5H_{10}$ |
| 4-Hydroxy-3-Methoxybenzaldehyde (Vanillin) | $C_8H_8O_3$ |
| Napthalene | $C_{10}H_8$ |
| Pyridine | $C_5H_5N$ |
| Pyrimidine | $C_4H_4N_2$ |
| 7H-purin-6-amine (Adenine) | $C_5H_5N_5$ |
| 5-methyl-1H-pyrimidine-2,4-dione (Thymine) | $C_5H_6N_2O_2$ |
| 6-amino-1H-pyrimidin-2-one (Cytosine) | $C_4H_5N_3O$ |
| 2-amino-1,7-dihydropurin-6-one (Guanine) | $C_5H_5N_5O$ |

Table 3 List of molecules and their chemical formula for which predictions were made from a network trained with the 25 datasets from Table 1 without the ionization potential as an input parameter.

For both the alkanes and alkenes, the predicted cross sections increased with the size of the molecule, a trend that was also observed in experimental cross section data for the molecules Ethanal, Propanal, and Butanal; Methanol and Ethanol; and Propanone and Butanone (see Fig. 1). Comparing the predicted cross sections for the alkanes to the alkenes shows that either the double bond or the reduced number of hydrogen atoms in the molecule in the alkenes lowers the cross sections relative to the alkanes. For molecules with a ring structure, the predicted cross sections again increase in magnitude as the number of atoms in the molecule increases. For the nucleotide bases, the cross section predictions for Adenine and Thymine are nearly identical, which is unexpected given the differences in the chemical formulae. Since Adenine is the only molecule without an oxygen atom, one might expect that its cross section would differ from the other nucleotide bases.

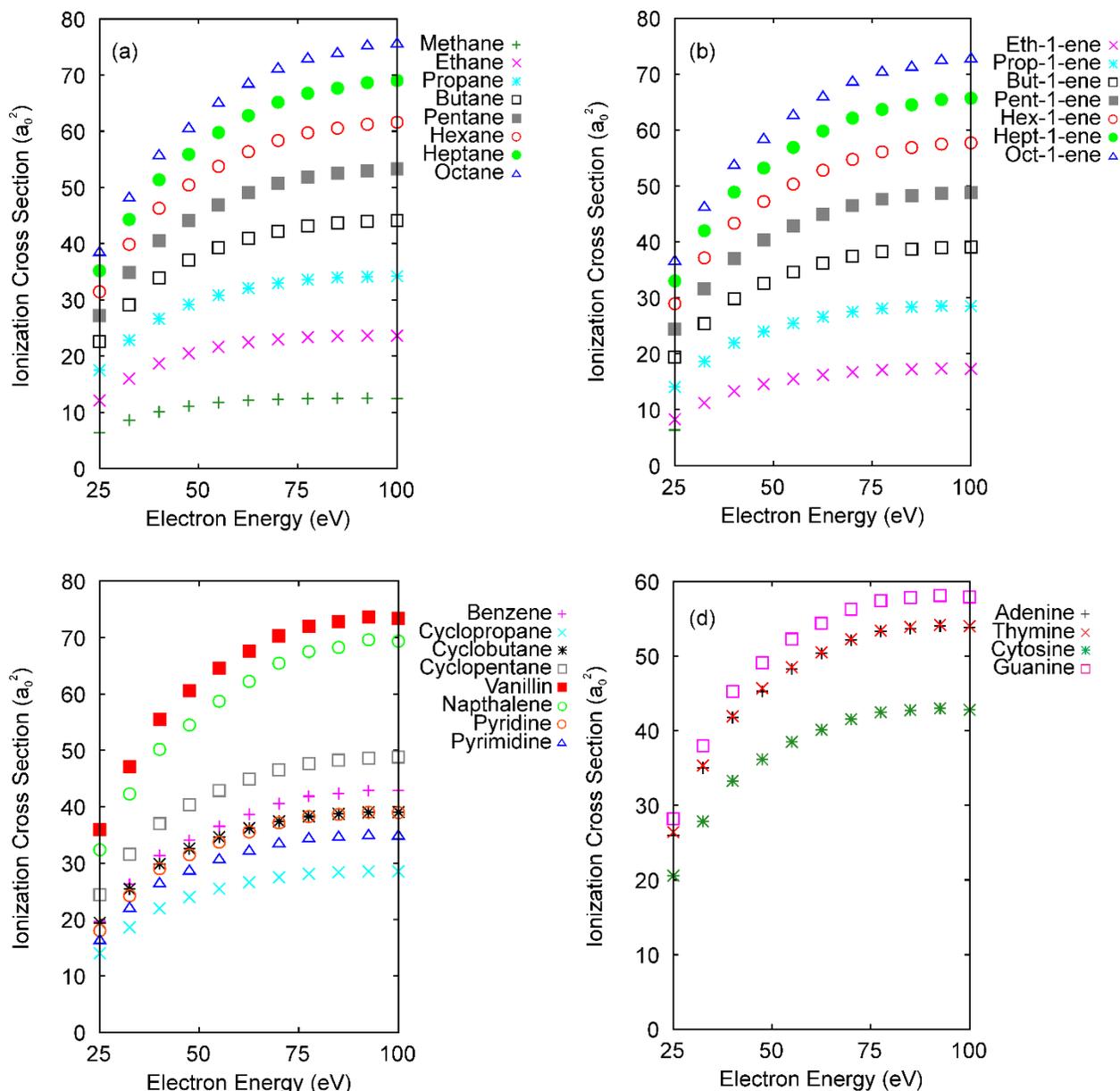

Figure 6 Predictions for ionization cross sections for new molecules using a network trained on all 25 input datasets listed in Table 1. (a) alkanes, (b) alkenes, (c) molecules with a ring structure, (d) nucleotide bases.

## 4. Conclusions

Due to their nuclear and electronic complexity, the calculation or measurement of collision cross sections for molecular targets remains a challenging task for theory and experiment. We have introduced a machine learning model capable of predicting ionization cross sections for a wide array of molecular targets with limited training data and input parameters. We used 3-layer feed-forward neural networks with inputs of the number of carbon, hydrogen, nitrogen, and oxygen atoms in the molecule, as well as the ionization potential, and outputs of the molecular cross section at projectile energies between 25 and 100 eV. The networks were trained using published experimental data, making them independent of any approximations required in theoretical models and ensuring that all physics captured in the experiment is included in the trained network.

Overall, our results showed that networks that used the chemical formula of the molecule as inputs, but not the ionization potential, resulted in the most accurate predictions for the cross sections, with the prediction differing from the known experimental result by at most 30% and frequently less than 10%. This indicated that there is a strong correlation between the chemical formula of the molecule and its cross section, but a weak correlation between the ionization potential and the cross section. The strong correlation of cross section with chemical formula is something that has been exploited for many decades when trying to predict molecular ionization cross sections. For example, one of the simplest models for predicting the cross sections is the Bragg additivity rule, in which the molecular cross section is estimated as the sum of the atomic cross sections for all the atoms in the molecule. A number of variations of this additivity rule have been introduced, including using a weighted sum of atomic cross sections [30], the Screening Corrected Additivity Rule (SCAR) that accounts for some geometric overlap between the constituent atoms [31], and the Pixel Counting Method (PCM) [32] that includes molecular orientation and atomic overlap effects. All of these methods are based on the idea that the molecular cross section can be expressed as some function of the atomic cross sections, and our ML model exploits a similar relationship.

Expectedly, networks that included more training data yielded more accurate predictions. However, even with as few as 10 training datasets, the network was able to predict the cross section to within a reasonable degree of accuracy. The identity of the molecules included in the training set had an effect on the success of the predictions, but this variation was again within acceptable limits. The uncertainty in our model predictions was typically similar to experimental uncertainties.

By using a network trained on 25 published molecular cross section datasets, we made predictions for a wide variety of molecules, including alkanes, alkenes, molecules with ring structures, and DNA nucleotide bases. These predicted cross sections exhibited the trends that molecules with more atoms had a larger cross section and molecules with similar chemical formulae yielded similar cross sections. These trends were consistent with published experimental data.

Prior ML models for predicting cross sections have so far either been limited to a small number of target molecules or relied on theoretical models to provide training data. The algorithm introduced here is broadly applicable to any molecule whose constituent atoms are carbon, hydrogen, nitrogen, or oxygen and is not limited by theoretical approximations. Training of a given network required less than 10 minutes on a laptop computer and predictions from a fully trained network require only a fraction of a second. Thus, our trained network offers fast and accurate predictions of molecular ionization cross sections. We have also made the algorithm available open access so that other researchers may use it for molecular targets beyond those we have examined.

It is possible that further improvements to the algorithm could be made by changing the network structure, using a different ML technique, including additional input parameters, expanding the applicable energy range, or increasing the possible constituent atoms. However, these investigations are left to future studies. The success demonstrated by the current model provides additional evidence to a growing body of work that demonstrates the value of ML models for estimating collision cross sections. These techniques will hopefully continue to provide a quick and accurate source of approximate cross section data for use in many applications.

## Acknowledgements
We gratefully acknowledge the support of the National Science Foundation under Grant No. PHY-1912093.

## References

[1] C. Powell, *NIST Electron Elastic-Scattering Cross-Section Database, NIST Standard Reference Database 64*, https://doi.org/10.18434/T4NK50.

[2] LxCat Database, *LxCat*, https://nl.lxcat.net/home/.

[3] BEAMDB, *BEAMDB - Belgrade Electron/Atom(Molecule) Database*, http://servo.aob.rs/emol/.

[4] VAMDC, *Databases | VAMDC Consortium*, http://www.vamdc.org/structure/databases/.

[5] *Atomic, Molecular, and Optical Science Gateway*, (unpublished).

[6] D. Guest, K. Cranmer, and D. Whiteson, *Deep Learning and Its Application to LHC Physics*, Annu. Rev. Nucl. Part. Sci. **68**, 161 (2018).

[7] A. Radovic, M. Williams, D. Rousseau, M. Kagan, D. Bonacorsi, A. Himmel, A. Aurisano, K. Terao, and T. Wongjirad, *Machine Learning at the Energy and Intensity Frontiers of Particle Physics*, Nature **560**, 7716 (2018).

[8] G. Carleo and M. Troyer, *Solving the Quantum Many-Body Problem with Artificial Neural Networks*, Science **355**, 602 (2017).

[9] V. Dunjko and H. J. Briegel, *Machine Learning & Artificial Intelligence in the Quantum Domain: A Review of Recent Progress*, Rep. Prog. Phys. **81**, 074001 (2018).

[10] K. T. Butler, D. W. Davies, H. Cartwright, O. Isayev, and A. Walsh, *Machine Learning for Molecular and Materials Science*, Nature **559**, 7715 (2018).

[11] Salah Yaseen El-Bakry and Mahmoud Yaseen El-Bakry, *Neural Network Representation for Electron and Positron Collisions with Sodium and Potassium Atoms*, Indian J. Phys **78**, 1313 (2004).

[12] A. L. Harris and J. A. Darsey, *Applications of Artificial Neural Networks to Proton-Impact Ionization Double Differential Cross Sections*, The European Physical Journal D **67**, (2013).

[13] S. Y. El-Bakry, E.-S. El-Dahshan, and M. Y. El-Bakry, *Total Cross Section Prediction of the Collisions of Positrons and Electrons with Alkali Atoms Using Gradient Tree Boosting*, Indian J Phys **85**, 1405 (2011).

[14] L. Zhong, *Fast Prediction of Electron-Impact Ionization Cross Sections of Large Molecules via Machine Learning*, Journal of Applied Physics **125**, 183302 (2019).

[15] P. H. R. Amaral and J. R. Mohallem, *Machine-Learning Predictions of Positron Binding to Molecules*, Phys. Rev. A **102**, 052808 (2020).

[16] A. Jasinski, J. Montaner, R. C. Forrey, B. H. Yang, P. C. Stancil, N. Balakrishnan, J. Dai, R. A. Vargas-Hernández, and R. V. Krems, *Machine Learning Corrected Quantum Dynamics Calculations*, Phys. Rev. Res. **2**, 032051 (2020).

[17] V. Jetly and B. Chaudhury, *Extracting Electron Scattering Cross Sections from Swarm Data Using Deep Neural Networks*, Mach. Learn.: Sci. Technol. **2**, 035025 (2021).

[18] P. W. Stokes, R. D. White, L. Campbell, and M. J. Brunger, *Toward a Complete and Comprehensive Cross Section Database for Electron Scattering from NO Using Machine Learning*, The Journal of Chemical Physics **155**, 084305 (2021).

[19] P. W. Stokes, D. G. Cocks, M. J. Brunger, and R. D. White, *Determining Cross Sections from Transport Coefficients Using Deep Neural Networks*, Plasma Sources Sci. Technol. **29**, 055009 (2020).
[20] A. L. Harris and J. Nepomuceno, *Predict Molecular Cross Sections (Pmx.Nb)*, (2023) doi: 10.6084/m9.figshare.24082035.
[21] L. Fausett, *Fundamentals of Neural Networks: Architectures, Algorithms, and Applications* (Prentice Hall, Englewood Cliffs, 1994).
[22] Wolfram, *Mathematica*, (2022).
[23] J. N. Bull and P. W. Harland, *Absolute Electron Impact Ionization Cross-Sections and Polarisability Volumes for C2 to C4 Aldehydes, C4 and C6 Symmetric Ethers and C3 to C6 Ketones*, International Journal of Mass Spectrometry **273**, 53 (2008).
[24] D. Rapp and P. Englander-Golden, *Total Cross Sections for Ionization and Attachment in Gases by Electron Impact. I. Positive Ionization*, The Journal of Chemical Physics **43**, 1464 (1965).
[25] K. L. Nixon, W. A. D. Pires, R. F. C. Neves, H. V. Duque, D. B. Jones, M. J. Brunger, and M. C. A. Lopes, *Electron Impact Ionisation and Fragmentation of Methanol and Ethanol*, International Journal of Mass Spectrometry **404**, 48 (2016).
[26] M. Terrissol, M. C. Bordage, V. Caudrelier, and P. Segur, *Cross-Sections for 0.025 EV-1 KeV Electrons and 10 EV-1 KeV Photons*, Atomic and Molecular Data for Radiotherapy (IAEA-TECDOC-506), IAEA, Vienna 218 (1989).
[27] N. O. of D. and Informatics, *Welcome to the NIST WebBook*, https://doi.org/10.18434/T4D303.
[28] C. Q. Jiao, C. A. DeJoseph, R. Lee, and A. Garscadden, *Kinetics of Electron Impact Ionization and Ion-Molecule Reactions of Pyridine*, International Journal of Mass Spectrometry **257**, 34 (2006).
[29] I. Linert, M. Dampc, B. Mielewska, and M. Zubek, *Cross Sections for Ionization and Ionic Fragmentation of Pyrimidine Molecules by Electron Collisions*, Eur. Phys. J. D **66**, 20 (2012).
[30] C. Champion, H. Lekadir, M. E. Galassi, O. Fojón, R. D. Rivarola, and J. Hanssen, *Theoretical Predictions for Ionization Cross Sections of DNA Nucleobases Impacted by Light Ions*, Phys. Med. Biol. **55**, 6053 (2010).
[31] F. Blanco and G. García, *Screening Corrections for Calculation of Electron Scattering from Polyatomic Molecules*, Physics Letters A **317**, 458 (2003).
[32] H. J. Lüdde, A. Achenbach, T. Kalkbrenner, H.-C. Jankowiak, and T. Kirchner, *An Independent-Atom-Model Description of Ion-Molecule Collisions Including Geometric Screening Corrections*, Eur. Phys. J. D **70**, 82 (2016).